\documentclass[11pt,titlepage]{article}
\usepackage{amsmath,amsthm,amssymb,graphicx}
\usepackage[letterpaper,margin=1.25in]{geometry}

\usepackage{sectsty}
\subsectionfont{\it}

\usepackage{apacite}

\def\Wbar{\overline{W}}
\def\wbar{\overline{w}}
\newcommand{\GD}{\Delta}
\newcommand{\D}{\Delta}
\newcommand{\dd}{{\hbox{\rm d}}}
\newcommand{\Eq}[1]{Eq.~(\ref{eq:#1})}
\newcommand{\Gth}{\theta}
\title{Wright's adaptive landscape versus Fisher's fundamental theorem}

\author{Steven A.\ Frank\footnote{Department of Ecology and Evolutionary Biology, University of California, Irvine, CA 92697--2525, USA, and Santa Fe Institute, 1399 Hyde Park Road, Santa Fe, NM 87501, USA, email: safrank@uci.edu.\hfil\break \hbox{\null}\hskip10pt To appear as: Frank, S. A. 2011. Wright's adaptive landscape versus Fisher's fundamental theorem. Pages 000--000 in {\it The Adaptive Landscape in Evolutionary Biology.} E. Svensson and R. Calsbeek, eds. Oxford University Press.}}

\begin{document}

%% apacite modifications
%% special renew commands to get style close to JEB formatting

\renewcommand\BOthers{\emph{et al}\hbox{}}
\renewcommand\BOthersPeriod{\emph{et al}\hbox{}}
\renewcommand{\BCBT}{}				% ref list no final comma w/two authors 
\renewcommand{\BCBL}{}				% ref list no final comma w/more authors
\renewcommand{\APACrefYearMonthDay}[3]{#1}	% no parens around year in reflist
\renewcommand{\APACrefYear}[1]{#1}	% no parens around year in reflist
\renewcommand{\BEd}{edn\hbox{}}         	% edition
\renewcommand{\APACrefYear}[1]{%
  {#1}%
}
\renewcommand{\APACjournalVolNumPages}[4]{%
  \Bem{#1}%             journal
  \ifx\@empty#2\@empty
  \else
    \unskip\ \textbf{#2:}%  volume
  \fi
  \ifx\@empty#3\@empty
  \else
    \unskip%      issue number
  \fi
  \ifx\@empty#4\@empty
  \else
    \unskip\ {#4}%      pages
  \fi
}
\renewcommand{\APACaddressPublisher}[2]{%
  \ifx\@empty#1\@empty
    \ifx\@empty#2\@empty
    \else
      {#1}%                 address
    \fi
  \else
    {#2}%                   publisher
    \ifx\@empty#2\@empty
    \else
      \unskip, {#1}%        address
    \fi
  \fi
}
		% must put after begin doc

\maketitle

\@titlepagetrue
\if@titlepage\newpage\else\noterule\fi

\begin{abstract}
Two giants of evolutionary theory, Sewall Wright and R. A. Fisher, fought bitterly for over thirty years.  The Wright-Fisher controversy forms a cornerstone of the history and philosophy of biology.  I argue that the standard interpretations of the Wright-Fisher controversy do not accurately represent the ideas and arguments of these two key historical figures.  The usual account contrasts the major slogans attached to each name: Wright's adaptive landscape and shifting balance theory of evolution versus Fisher's fundamental theorem of natural selection.  These alternative theories are in fact incommensurable.  Wright's theory is a detailed dynamical model of evolutionary change in actual populations.  Fisher's theory is an abstract invariance and conservation law that, like all physical laws, captures essential features of a system but does not account for all aspects of dynamics in real examples.  This key contrast between embodied theories of real cases and abstract laws is missing from prior analyses of Wright versus Fisher.  They never argued about this contrast.  Instead, the issue at stake in their arguments concerned the actual dynamics of real populations.  Both agreed that fluctuations of nonadditive (epistatic) gene combinations play a central role in evolution.  Wright emphasized stochastic fluctuations of gene combinations in small, isolated populations.  By contrast, Fisher believed that fluctuating selection in large populations was the main cause of fluctuation in nonadditive gene combinations.  Close reading shows that widely cited views attributed to Fisher mostly come from what Wright said about Fisher, whereas Fisher's own writings clearly do not support such views.
\if@titlepage\vfill\setcounter{page}{2}\thispagestyle{plain}\fi

\end{abstract}
\if@titlepage\newpage\else\noterule\vfill\eject\fi
\if@titlepage\setcounter{page}{3}\fi

\begin{quote}
\textit{Fisher is frequently portrayed in the contemporary literature as believing in a strictly additive basis for the inheritance of quantitative characters, and as dismissing any evolutionary importance for epistatic interactions in fitness effects. This is accompanied by a sub-text that this is in some way less virtuous than embracing a less `reductionist'	view, which	assigns	a prominent	role to epistasis, as in	Wright's `shifting-balance' theory. This is, in fact, a travesty of Fisher's views \cite{charlesworth00book}.
}
\end{quote}

\section*{Introduction}

Wright developed the adaptive landscape to support his shifting balance theory of evolution.  The shifting balance theory emphasized that progressive improvement by natural selection is too slow by itself to account for biological diversity and the rate of adaptive change.  Wright suggested that random perturbations of gene frequencies in small partially isolated populations may act synergistically with natural selection to explain rapid adaptation and diversity. One may visualize the synergy of random perturbations and deterministic natural selection by imagining the dynamics of populations on an adaptive landscape.

In the adaptive landscape, natural selection corresponds to climbing local hills of increasing fitness.  A local peak traps a population to a narrow range of phenotypes that limits opportunities for major improvement in fitness.  However, a small population may, by stochastic sampling and drift, change its common nonadditive (epistatic) combinations of interacting genes.  Such perturbations of epistatic gene combinations can move a population down a hill and across a valley of lower fitness to the base of a nearby and potentially higher fitness peak.  Natural selection then pushes the population up that higher peak, causing a major improvement in fitness relative to the recently abandoned lower peak.  Such peak shifts lead to major diversification of phenotype---the shifting balance process { (Chapter 5, Wade, this volume)}.  

Wright often presented the local hill climbing aspect of natural selection in terms of Fisher's fundamental theorem.  That theorem describes the rate of improvement in fitness caused by natural selection.  Wright also ascribed to Fisher the view that local hill climbing by natural selection was the primary force of evolutionary change over long periods of time.  Such local hill climbing does not require a key role for nonadditive gene combinations, so Wright also ascribed to Fisher the view of natural selection acting locally on additive gene effects { (Chapter 6, Goodnight, this volume)}.

Fisher strongly rejected Wright's characterization of the fundamental theorem and, in turn, severely criticized the adaptive landscape.  At first glance, it may seem that the Wright-Fisher controversy ultimately comes down to the opposing views given by each combatant's primary slogan:  the adaptive landscape on Wright's side versus the fundamental theorem on Fisher's side.

I clarify two points.  First, Wright and Fisher did disagree about whether random perturbations of drift were essential to explain the long-term processes of adaptation and diversification.  That disagreement was in fact their primary battle.  Much of their sometimes acrimonious mischaracterizations of each other's work on various topics often derived from this single and often unspoken rift with regard to the importance of stochastic fluctuations in small populations.  They did not disagree about whether such fluctuations occurred, only the relative importance of those fluctuations in adaptive evolution \cite{provine86sewall}.

My second point concerns the very different goals of Wright and Fisher with respect to the adaptive landscape and the fundamental theorem.  Wright spent decades of intensive work refining the adaptive landscape theory.  He made that effort to provide support for the shifting balance theory as the prime mover of evolutionary change and biological diversity.  

Fisher presented the fundamental theorem in his 1930 book and rarely commented on it again except to criticize Wright or provide a few minor corrections.  Wright actually produced more commentary on the fundamental theorem than Fisher.  However, Wright's commentary almost always misrepresents both Fisher's particular results and Fisher's deeper goals for the fundamental theorem.  When discussing Fisher's work, Wright promoted his own views as defining the long term consequences of nonadditive gene interactions over a global multipeaked landscape and Fisher's views as defining the short term consequences of additive gene action within a local and narrowly confined fitness peak.  

Why did Fisher let Wright's misrepresentations go mostly unanswered?  In my opinion, Fisher did not see the fundamental theorem as having anything to do with their primary disagreement over the roles of geographic isolation and random perturbations in long term adaptive evolution.  The fundamental theorem is about the logical nature of selection as a universal law of biology.  That law expresses an invariant rate of change caused by natural selection when considered alone as an isolated force, as distinct from the total change to a population caused by a variety of processes including mutation, recombination, competition, and so on.  Fisher also presented a quasi-conservation principle: the amount of adaptive improvement by natural selection is typically balanced by an equal and opposite decline in fitness caused by increased competition from simultaneously improved competitors.  The total fitness of a population must typically change hardly at all, because population growth rates must typically be close to zero.  A continuously growing population would overrun the world; a continuously declining population would soon be extinct.  

Fisher had training in mathematical physics.  For him, the fundamental theorem had the same power in biology that the great laws of invariance and conservation had in physics.  Those physical laws do not predict how a real, complex, heterogeneous and open physical system will evolve over time.  Such predictions of complex dynamics in real systems are often impossible and at best not reducible to a brief and simple expression.  In the same way, Fisher never suggested that the fundamental theorem predicted how real populations would evolve over time. Rather, he intended only to express how natural selection as a force necessarily acted within a complex evolutionary system subject to many distinct types of forces.  When Fisher and Wright argued, the issues primarily concerned how real populations evolve over long periods of time.  Fisher was very interested in the problem of long term evolution, but he also realized that the fundamental theorem had little to say on that topic.  

Wright's published commentary continues to define the dominant view of Fisher's outlook on evolutionary dynamics and on the fundamental theorem.  Here, I describe what Fisher actually wrote about these topics, which differs greatly from the picture painted by Wright.  

\section*{The disagreement about drift and dynamics}

In \citeA{fisher50the-sewall}, Fisher clearly expressed his deepest disagreement with Wright:
\begin{quote}
The widest disparity$\ldots$which has so far developed in the field of Population Genetics is that which separates those who accept from those who reject the theory of ``drift'' or ``non-adaptive radiation,'' as it has been called by its author, Professor Sewall Wright of Chicago.

[T]his theory of Sewall Wright$\ldots$claims that the subdivision of a population into small isolated or semi-isolated colonies has had important evolutionary effects; and this through the agency of random fluctuation of gene ratios, due to random reproduction in a small population.

We have long felt that there are grave objections to this view$\ldots$[O]ne, however, is completely fatal to the theory in question, namely that it is not only small isolated populations, but also large populations, that experience fluctuations in gene ratio. If this is the case, whatever other results isolation into small communities may have, any effects which flow from fluctuating variability in the gene ratios will not be confined to such subdivided species, but will be experienced also by species having continuous populations.

This fact, fatal to ``The Sewall Wright Effect,'' appeared in our own researches from the discovery that the year-to-year changes in the gene ratio in a wild population were considerably greater than could be reasonably ascribed to random sampling, in a population of the size in question.
\end{quote}

Fisher and Ford agreed that random fluctuations by sampling and drift will always occur. But they argued that the fluctuations they observed were too great to be explained by sampling.  Instead, fluctuating selection caused by a varying environment appeared to be the cause.  They noted that others, such as Dobzhansky, have also presented data on fluctuating gene ratios (frequencies) most likely explained by selection.  Fisher and Ford conclude:

\begin{quote}
Sub-division into small isolated or semi-isolated populations is clearly favourable to evolutionary progress through the variety of environmental conditions to which the colonies are exposed. Moreover, so long as it could be believed that large fluctuations in gene ratios occur only in small isolated colonies by reason of fluctuations of random survival, then it might have been true that such fluctuations themselves favoured evolutionary change in a way that would not be allowed in a continuous distribution of the species. If now it is admitted that large populations with continuous distributions also show year-to-year fluctuations of comparable or greater magnitude in their gene ratios, due to variable selection, the situation is entirely altered. In these circumstances, the claim for ascribing a special evolutionary advantage to small isolated communities due to fluctuations in gene ratios, had better be dropped.
\end{quote}

Fisher and Ford are not saying that major adaptive changes occur only in large, panmictic populations.  Rather, they argue that subdivision into small populations and drift are not necessary conditions for significant adaptive change by natural selection.  The fluctuating gene frequencies in large populations caused by fluctuating selection may be sufficient to allow shifts in favored gene combinations and the equivalent of a Wrightian peak shift.  To repeat, Fisher's primary argument about major adaptive change is against a necessary role of subdivision, small population size, and drift.  Those factors may occur, but major adaptive shifts by altered epistatic gene combinations can arise in other ways.  

\citeA[pp.~301--302]{provine86sewall} clearly traces the origin of this disagreement between Fisher and Wright to the early 1930s.  At that time, the evolution of dominance formed the particular subject of debate, rather than the dynamics of gene frequencies under selection as in the Fisher and Ford paper above.  But, as Provine emphasizes, the real argument in the early 1930s that led to the original rift between Fisher and Wright also turned on the alternative views of selection and drift. Provine makes his case by quoting from \citeA[p.~50--51]{wright34physiological}:

\begin{quote}
From the standpoint of the theory of dominance it may seem of little importance which mechanism is accepted if it be granted that selection has been an important factor. This is not at all the case, however, with the implication of Fisher's and Plunkett's selection theories, for the theory of evolution. Fisher used the observed frequency of dominance as evidence for his conception of evolution as a process under complete control of selection pressure, however small the magnitude of the latter. 

My interest in his theory of dominance was based in part on the fact that I had reached a very different conception of evolution (1931) and one to which his theory of dominance seemed fatal if correct. As I saw it, selection could exercise only a loose control over the momentary evolutionary trend of populations. A large part of the differentiation of local races and even of species was held to be due to the cumulative effects of accidents of sampling in populations of limited size. Adaptive advance was attributed more to intergroup than intragroup selection.
\end{quote}

\citeA[pp.~302]{provine86sewall} nicely summarizes the key point:

\begin{quote}
I think Wright is correct in saying that what really was at stake in the argument with Fisher over the evolution of dominance was not the particular problem of dominance but their differing conceptions of evolution.  If either was correct on the evolution of dominance, it was perceived by the other as fatal to his entire conception of evolution.
\end{quote}

Fisher's criticism of the adaptive landscape focused on the claim that drift by random sampling is not a necessary condition for significant evolutionary change.  Wright's early work did specify random sampling as the key perturbation in small, local populations.  However, Wright expanded his framing of drift in later work, probably in response to Fisher and Ford's argument that fluctuating selection could explain how populations may be perturbed from a fixed, local peak on an adaptive landscape.  In Wright's \citeyear[p.~455]{wright77evolution} grand synthesis, he describes the first phase of the shifting balance process as:

\begin{quote}
\textit{Phase of Random Drift.} In each deme, the set of gene frequencies drifts at random in a multidimensional stochastic distribution about the equilibrium set characteristic of a particular fitness peak or goal.  The set of equilibrium values is the resultant of three sorts of pressures on the gene frequencies: those due to recurrent mutation, to recurrent immigration from other demes, and to selection.  The fluctuations in the gene frequencies responsible for the stochastic distribution (or random drift) may be due to accidents of sampling or to fluctuations in the coefficients measuring the various pressures [e.g., mutation, immigration, and selection].
\end{quote}

Here, Wright clearly allows that fluctuating selection may be the cause of perturbations to local populations.  Fisher was long dead by this time.  Fisher might have replied that fluctuating selection works just as well in large populations, so there would in this case be no need to invoke small separated populations as essential to the process.  Wright, in turn, may have answered that many small separated populations allow the many parallel independent lines an opportunity to initiate a peak shift, greatly increasing the chance that one local population makes the jump to another peak and then exports its enhanced adaptive combinations through the population.  In this view, Wright's primary idea is subdivision of the population into local populations, allowing multiple parallel exploration of the adaptive landscape and thereby greatly accelerating the pace of evolutionary change.  Fisher probably would have accepted that subdivision might under some conditions have an effect on evolutionary rate, but that such subdivision is neither necessary nor likely to be a commonly important factor.   

\section*{Fisher's goal for the fundamental theorem}

\citeA{fisher58the-genetical} stated the fundamental theorem as: ``The rate of increase in fitness of any organism at any time is equal to its genetic variance in fitness at that time'' (p.~37) and ``The rate of increase of fitness of any species is equal to its genetic variance in fitness'' (p.~50). At first glance, these expressions seem closely related to Wright's study of the adaptive landscape, which is usually described as a surface of population mean fitness.  Fisher's result would then describe how fast natural selection can push a population up a surface of mean fitness.  

Wright frequently quoted the fundamental theorem in support of his gradient formulation of the adaptive landscape, in which gene frequencies change at a rate proportional to the slope of the fitness surface, $\dd\Wbar/\dd q$, where $\Wbar$ is the mean fitness of the population, and $q$ is the frequency of a gene.  In what I believe to be Wright's \citeyear[p.~118]{wright88surfaces} last publication, he said: ``The effects [on gene frequencies in an adaptive landscape] may be calculated using Fisher's fundamental theorem.''

These quotes from both Fisher and Wright seem to say that Fisher's theorem is about the rate of change in the mean fitness of a population.  That interpretation was adopted by essentially everyone who subsequently commented on the theorem until papers by \citeA{price72fishers} and \citeA{ewens89an-interpretation} that I will come to later.  But we can see clearly from other statements by Fisher that something is wrong: ``In regard to selection theory, objection should be  taken to Wright's equation [the expression $\dd\Wbar/\dd q$] principally because it represents natural selection, which in reality acts upon individuals, as though it were governed by the average condition of the species or inter-breeding group'' \cite[p.~58]{fisher41average} and ``I have never, indeed, written about $\wbar$ and its relationships$\ldots$the existence of such a potential function [i.e. a function nondecreasing in time]$\ldots$is not a general property of natural populations, but arises only in the special and restricted cases which Wright has chosen to consider.'' \cite[p.~290]{fisher58polymorphism}.

Fisher and Wright never suggested the other's equations were incorrect. Once again the disagreement is about how to interpret evolutionary process.  Wright's goal remains easy to follow. He wanted to understand the various forces that change gene frequency in order to argue for his shifting balance theory of evolution.  In developing his theory, he needed expressions for how natural selection changes gene frequencies.  Wright repeatedly invoked Fisher's fundamental theorem to describe how natural selection moves populations up a hill of increasing fitness.  By contrast, Fisher's goal for the fundamental theorem seems obscure at first glance.

When Fisher argued against Wright's shifting balance theory, he clearly focused on the key issues of population subdivision and the role of drift in perturbing gene frequencies in small, isolated populations.  Thus, Fisher's complaint about Wright's use of the fundamental theorem does not have to do with shifting balance and the controversy over long term evolutionary dynamics.  If not about shifting balance and evolutionary dynamics, what was Fisher ultimately arguing by saying the fundamental theorem expressed ``the rate of increase in fitness of any species'' and at the same time sharply criticizing Wright by saying ``In regard to selection theory, objection should be  taken to Wright's equation$\ldots$principally because it represents natural selection$\ldots$as though it were governed by the average condition of the species'' and ``I have never, indeed, written about $\wbar$ [mean fitness] and its relationships''?

\subsection*{Background}

Essentially everyone interpreted Fisher's theorem in relation to the long term dynamics of populations.  The theorem seemed to say, at the very least, that the average fitness of a population never decreased.  More strongly, the theorem described the dynamical path of mean fitness in relation to genetic variance.  

\citeA{fisher30the-genetical,fisher58the-genetical} emphasized strongly that his theorem is exact.  Yet essentially every commentator in the forty years following the 1930 announcement qualified the theorem by the wide variety of special assumptions required: random mating, large populations, pure additivity of genic interactions (no epistasis), free recombination with no linkage disequilibrium, and no frequency or density dependent interactions.  Several analyses showed that mean fitness could decrease under a variety of conditions.  Other analyses quantified how closely mean fitness tracked additive genetic variance and thus the extent to which the fundamental theorem was a good approximate result under certain special conditions.  

\citeA{price72fishers} provided the first clues about the theorem as Fisher meant it.  \citeA{ewens89an-interpretation} followed with a full, clear proof and exposition. The Price-Ewens exposition showed that Fisher never meant to discuss the long term dynamics of populations.  Thus, Wright's use of the theorem and all of the prior commentary about evolutionary dynamics had nothing to do with Fisher's view of the theorem.

I do not give the mathematical details here.  Interested readers should consult the extensive literature that has developed, which can be found by tracing citations to \cite{price72fishers} and \cite{ewens89an-interpretation}.  My own more technical interpretations are in \citeA{frank97the-price,frank09natural}.  Here, I give a simplified expression of the key ideas based on \citeA{frank92fishers}.

\subsection*{Fisher's framing of the problem}

Fisher realized that one cannot make a complete model of evolutionary dynamics.  Too many factors come into play: changes in the physical environment, changes in competitive intensity within and between species, and changes in the complex nonadditive interactions between genes that fluctuate in frequency.  Given the complexity of ``open'' systems in which forces flow from a variety of unknown sources, Fisher sought a way to define a ``closed'' subset in which one could completely and exactly study the process of natural selection. Indeed, the first sentences of \textit{The Genetical Theory\/} are \cite{fisher58the-genetical}:

\begin{quote}
Natural Selection is not Evolution.	Yet, ever since	the	two words have been in common use, the theory of Natural Selection has been employed as a convenient abbreviation for the theory of Evolution by means of Natural Selection, put forward by Darwin and Wallace. This has had the unfortunate consequence that the theory of Natural Selection itself has scarcely ever, if ever, received separate consideration. To draw a physical analogy, the laws of conduction of heat in solids might be deduced from the principles of statistical mechanics, yet it would have been an unfortunate limitation, involving probably a great deal of confusion, if statistical mechanics had only received consideration in connexion with the conduction of heat. In this case it is clear that the particular physical phenomena examined are of little theoretical interest compared to the principle by which they can be elucidated. The overwhelming importance of evolution to the biological sciences partly explains why the theory of Natural Selection should have been so fully identified with its role as an evolutionary agency, as to have suffered neglect as an independent principle worthy of scientific study.
\end{quote}

The expression of intent seems clear. Fisher wishes to isolate natural selection as a process from the context of particular aspects evolutionary dynamics as they occur in particular instances.  Put another way, to study evolutionary dynamics, one must make many particular assumptions that confine the analysis to a particular kind of problem, and so obscure any general principles that may hold for natural selection across all assumptions and particular instances of evolutionary dynamics.

\subsection*{The fundamental theorem explained}

Fisher started his argument by first isolating the general aspects of natural section from those aspects of evolutionary dynamics that are particular to each system.  To do this, Fisher set the standard for measurement of fitness as the full conditions of the population and environment at a particular instant in time.  

Those conditions, together called ``environment,'' include all of the gene frequencies that set the genetic environment in which each gene lives, all of the biotic interactions within and between species, and all aspects of the physical environment.  By fixing those environmental conditions at a particular instant, Fisher obtained a fixed standard against which he could measure the exact contribution of natural selection to changes in the adaptation of populations.  Fisher fully recognized that the actual evolutionary change in adaptation and mean fitness would then include two components: one component caused by natural selection in relation to the original fixed environmental standard of measurement, and one component caused by the changes in the environmental standard of measurement.  

Perhaps the most confusing aspect arises because natural selection itself changes the environmental standard of measurement by changing gene frequencies (genetic environment), by changing competitive intensity, and perhaps by changing the physical environment.  Those effects of natural selection on the standard of measurement are not, in Fisher's system, direct components ascribed to natural selection, but rather indirect components that Fisher lumped into the term for changes in the environment.  Although such a partitioning of total evolutionary change may seem arbitrary with regard to defining the consequences of natural selection, there is no other way to isolate the role of natural selection, because natural selection is a force that acts instantaneously in relation to the conditions that hold at that instant.  Once one sees this point of view, all else is detail.

\citeA{frank92fishers} expressed Fisher's partition as follows.  The total change in fitness over time, $\D\Wbar$, in the context of the environment, $E$, can be defined as
\begin{equation*}
\D\Wbar = \Wbar'|E' - \Wbar|E,
\end{equation*}
where $\Wbar|E$ is mean fitness in the context of a particular environmental state, primes denote one time step or instant into the future, and $\D\Wbar$ is the total change in fitness which nearly everyone had assumed was the object of Fisher's analysis. Fisher's theorem, however, was not concerned with the total evolutionary change, which depends at least as much on changes in the environment as it does on natural selection. Instead, Fisher partitioned the total change into
\begin{eqnarray}
\D\Wbar &=& \left(\Wbar'|E - \Wbar|E\right) +\left(\Wbar'|E' - \Wbar'|E\right)\notag\\
        &=& \D\Wbar_{NS} + \D\Wbar_E.\label{eq:partition}
\end{eqnarray}

Fisher called the first term the change in fitness caused by natural selection because there is a constant frame of reference, the initial environmental state, $E$. The fundamental theorem proves that the change in fitness caused by natural selection is equal to the genetic variance in fitness, where genetic variance is defined in a particular way (see below). \citeA[p.~45--46]{fisher58the-genetical} referred to the second term as the change caused by the environment, or as the change caused by the deterioration of the environment to stress that this term is often negative, because natural selection increases fitness but the total change in fitness is usually close to zero:  

\begin{quote}
Against the action of Natural Selection in constantly increasing the fitness of every organism, at a rate equal to the genetic variance in fitness which that population maintains, is to be set off the very considerable item of the deterioration of its inorganic and organic environment.  This at least is the conclusion which follows from the view that organisms are very highly adapted.  Alternatively, we may infer that the organic world in general must tend to acquire just that level of adaptation at which the deterioration of the environment is in some species greater, though in some less, than the rate of improvement by Natural Selection, so as to maintain the general level of adaptation nearly constant$\ldots$ 

An increase in numbers of any organism will impair its environment in a manner analogous to, and more surely than, an increase in the numbers or efficiency of its competitors. It is a patent oversimplification to assert that the environment determines the numbers of each sort of organism which it will support.  The numbers must indeed be determined by the elastic quality of the resistance offered to increase in numbers, so that life is made somewhat harder to each individual when the population is larger, and easier when the population is smaller. The balance left over when from the rate of increase in the mean value of $m$ [fitness] produced by Natural Selection, is deducted the rate of decrease due to deterioration in environment, results not in an increase in the average value of $m$, for this average value cannot greatly exceed zero, but principally in a steady increase in population.

\end{quote}

\subsection*{Fisher's conservation law for mean fitness}

Fisher's argument that mean population growth rate (fitness) must always remain close to zero leads to an approximate conservation law: any increase in the mean fitness of a population caused by natural selection must usually be balanced by an equal and opposite decrease in mean fitness caused by ``deterioration of the environment.''  Here, deterioration would most often arise from increased competition by members of the same or different species, as those competitors also increase their own level of adaptedness by natural selection { (Chapter 7, Calsbeek et al., this volume)}.  

Fisher supported this approximate conservation law of mean fitness by arguing that total population growth cannot be continually above zero, otherwise the population would grow without bound.  Similarly, total population growth cannot be continually less than zero, otherwise the population would soon disappear.  Fisher recognized that one species can increase at the expense of other species, so the total mean growth rate applies to all species potentially in competition with each other.  

Fisher clearly emphasized this balance between improvement by natural selection and deterioration by enhanced competition.  However, this broad context of the theorem has been almost entirely ignored. Instead, the focus has been on the natural selection component of increase, as in the quote ``The rate of increase in fitness of any organism at any time is equal to its genetic variance in fitness at that time'' \cite[p.~50]{fisher58the-genetical}.  Wright's use and commentary of the theorem concerned only this first component.  So it is useful to look explicitly at Fisher's expression for the natural selection component of evolutionary change in mean fitness. 

\subsection*{Average excess, average effect, and genetic variance}

The fundamental theorem's logic and its relations to Wright's adaptive landscape depend on two key definitions.  Each definition quantifies the contribution of a particular allele to a character, in this case fitness.  Here, I give rough verbal descriptions to emphasize the main ideas.  Details can be found in \citeA{ewens89an-interpretation} and \citeA{frank97the-price}.  Note that minor variants of the definitions exist in the literature, but all forms have the same essential meaning. 

It is easiest to think of a single diploid genetic locus with two alleles, $B$ and $b$, and three genotypes, $BB$, $Bb$, and $bb$.  The \textit{average excess\/} measures the excess reproduction of $B$ relative to an average individual.  To calculate the excess, we start with the fitness of individuals with the $BB$ genotype and one-half of the fitness of individuals with the $Bb$ genotype, the half arising because the heterozygote carries half as many copies of $B$ as the homozygote.  From the average fitness for $B$ we subtract the average fitness of all individuals, leaving the excess reproduction of the $B$ allele compared with the population as a whole.  

The average excess is a direct measure of the change in gene frequency, because it simply counts up the number of newly made alleles of a particular type compared with the average number of newly made alleles in the population.  It is helpful to show this change in gene frequency in symbols. Suppose that each allele over all loci is associated with an index label $j$, with frequency $q_j$ and average excess $a_j$ in a population with average fitness $\Wbar$.  Then the change in the frequency of each allele after a round of reproduction is 
\begin{equation}\label{eq:aveExcess}
\D q_j = q_ja_j/\Wbar.
\end{equation}

\citeA[p.~31]{fisher58the-genetical} emphasized that the average excess is not a good measure of the direct contribution of an allele to fitness, but rather is defined simply to describe the change in gene frequency that arises from the distribution of fitnesses among genotypes:

\begin{quote}
The [average] excess in a factor will usually be influenced by the actual frequency$\ldots$ of the alternative genes, and may also be influenced, by way of departures from random mating, by the varying reactions of the factor in question with other factors.
\end{quote}

The \textit{average effect\/} is a more subtle measure of the contribution of a particular allele to fitness.  Take a population in its current form, fully accounting for nonrandom mating, linkage associations between loci, nonadditive epistatic interactions between genes, and so on.  Measure the average effect of the allele $B$ by taking each individual in the population and changing, one at a time, each copy of $B$ to $b$, and measuring the effect of that change on fitness.  The average of each of those changes is the average effect of a gene substitution.  The advantage of this definition is that all aspects of mating pattern and interactions between genes are automatically accounted, because the change is made in each actual genetic combination that exists in the population.  The average effect of an allele is the partial regression coefficient of the presence of that allele on fitness.  

We use the symbol $b_j$ for the average effect of the $j$th allele on fitness \cite<using the notation and definitions of>{frank97the-price}.  Then we can write the total change in fitness caused by natural selection as
\begin{equation*}
\D\Wbar_{NS} = 2\sum_j(\D q_j)b_j,
\end{equation*}
where the two arises because we assume two alleles at each locus in a diploid genetic system.  This equation says that we can calculate the total change in fitness by natural selection by summing up each change in allele frequency, $\D q_j$, and weighting that change by the average effect of that allele on fitness, $b_j$.  This form provides the clearest expression of Fisher's fundamental theorem.  One can also show that this expression is equivalent to the variance in the average effects, which Fisher called the genetic variance in fitness \cite{ewens89an-interpretation}.  Thus, the change in fitness caused by natural selection is equal to the genetic variance in fitness.  

\subsection*{What is fitness?}

A key problem concerns the definition of fitness itself.  Fisher referred to fitness as a rate of increase, but he was vague about the precise definition of what is actually measured.  Fisher's vagueness in the conception of fitness caused confusion over the status of the fundamental theorem as a universally true mathematical theorem.  However, the technical details of how one might define fitness are not needed to understand the history and the main conceptual points about Fisher's theorem.  For those readers interested in this issue, I have added a brief Appendix.

\subsection*{Deterioration of the environment}

What about the change in fitness caused by the ``change in the environment'' as expressed by $\D\Wbar_E$ in \Eq{partition}?  We account for environmental changes by the changes in the average effects.  To obtain the changes in average effects, we recalculate the average effect of each allele in the changed population, including any changes in interactions between genes, changes in the array of genotypes caused by mating pattern, changes in competition between individuals, and changes in the physical environment.  \citeA{frank97the-price} wrote the total change in the environment as the total change in average effects
\begin{equation*}
\D\Wbar_E = 2\sum_j q_j'(\D b_j),
\end{equation*}
where the prime on $q$ shows that we use the frequencies in the changed populations to weight the changes in average effects for each allele.  Putting the pieces together and using the definitions in Equation (1) of \citeA{frank97the-price} yield the full partition 
\begin{eqnarray*}
\D\Wbar &=& \D\Wbar_{NS} + \D\Wbar_E\\
        &=& 2\sum_j(\D q_j)b_j+2\sum_j q_j'(\D b_j).
\end{eqnarray*}
The natural selection term is equivalent to the genetic variance in fitness. Conservation of total fitness implies that the deterioration of the environment term is typically close to the negative of the first term.  

Recently, I have shown that the genetic variance in fitness can also be thought of as a distance between the population before natural selection and after natural selection \cite{frank09natural}.  The distance measures the information the population acquires about the environment through the changes in gene frequencies caused by natural selection.  Quantifying the consequences of natural selection by an informational measure is conceptually more profound than quantifying the change in fitness by the genetic variance, although the descriptions are mathematically equivalent.  I will not pursue here my own informational interpretation, although that interpretation may be necessary to understand the full significance of the fundamental theorem as a law.

\subsection*{Misunderstandings about additivity}

Fisher's genetic variance is calculated by adding the contribution of each individual allele independently, leading to its common description as the additive genetic variance.  This description suggests that the additive genetic variance ignores dominance and genic interactions, instead assuming that each allele has a fixed contribution that can be taken independently and additively with respect to other alleles.  For example, \citeA[p.~353]{wright30the-genetical} noted in his review of \citeA{fisher30the-genetical}:

\begin{quote}
One's first impression is that the genetic variance in fitness must in general be large and that hence if the theorem is correct the rate of advance must be rapid. As Dr.\ Fisher insists, however, the statement must be considered in connection with the precise definition which he gives of the terms. He uses ``genetic variance'' in a special sense. It does not include all variability due to differences in genetic constitution of individuals. He assumes that each gene is assigned a constant value, measuring its contribution to the character of the individual (here fitness) in such a way that the sums of the contributions of all genes will equal as closely as possible the actual measures of the character in the individuals of the population. Obviously there could be exact agreement in all cases only if dominance and epistatic relationships were completely lacking. Actually, dominance is very common and with respect to such a character as fitness, it may safely be assumed that there are always important epistatic effects. Genes favorable in one combination, are, for example, extremely likely to be unfavorable in another. Thus allelomorphs which are held in equilibrium by a balance of opposing selection tendencies$\ldots$may contribute a great deal to the total genetically determined variance but not at all to the genetic variance in Fisher's special sense, since at equilibrium there is no difference in their contributions.
\end{quote}

This quote is the sort of commentary from Wright that led many people to regard Fisher's view as one of genes acting additively and ignoring Wright's own emphasis on the importance of dominance and epistatic genetic interactions.  However, one must parse this quote with care to understand what Wright is truly emphasizing.

The quote begins by framing the problem with respect to the rate of adaptive change.  Wright characterizes Fisher's argument as inevitably leading to the conclusion that the rate of adaptive change by natural selection must in fact be slow, because Fisher's analysis strips away the most important contributions to variance that come from nonadditive genetic interactions.  Wright continues by stressing the great importance of gene interactions, implying that a true theory of adaptive change must be based primarily on such interactions.  The quote is not really about Fisher's theorem, but rather about Wright's characterization of his difference with Fisher.  

According to Wright, the Wright view fully accounts for genetic interactions as the primary source of genetic variation and thus can fully account for the processes that may lead to rapid adaptive change.  By contrast, Wright has Fisher limited to the small component of genetic variance associated with the purely additive effects of genes acting in isolation, and thus with a theory that must be associated with a very limited rate of adaptive evolution.  

It is never quite true that Wright misunderstands Fisher's mathematics and arguments.  Wright understood mathematical genetics far too well for that.  But Wright's insistence on emphasizing his view of the Wright-Fisher contrast makes it very hard to get a fair characterization of Fisher's views from Wright.  Of course, Fisher did no better in return.  So, to understand their theories, we cannot read Wright on Fisher or read Fisher on Wright.

Fisher did not actually ignore dominance or genetic interactions.  Instead, he fully and completely accounted for those interactions.  The heritable contribution of each allele in the context of all of the genetic interactions in the population at any moment in time is exactly the average effect of the allele.  Fisher was trying to quantify the evolutionary change in fitness caused by natural selection, which means that the only important quantity with respect to each allele is its heritable contribution to fitness.  Heritable effects are the only effects that are passed to offspring, so they are the only effects that one must account in the calculation of change by natural selection. 

The average effect of each allele is chosen statistically to be the effect one has to add to an individual carrying the allele to get the best prediction of the individual's phenotype or fitness.  The average effect is a statistical form of additivity that accounts for all forms of nonadditive gene interactions. The average effect is not a physiological statement about the presence or absence of dominance or genetic interactions.  Hidden in Wright's statements is his own primary interest in how processes other than natural selection might rearrange the patterns of genetic interactions, thereby providing a different subsequent evolutionary path by natural selection. That sort of rearrangement of genetic interactions is a very interesting problem, but it has nothing to do with the fundamental theorem or with Fisher's accounting for additive and nonadditive genetic effects with respect to natural selection.

\section*{Fisher's laws versus Wright's dynamics}

The fundamental theorem expresses two laws.  First, the rate of increase in fitness caused by natural selection is an invariant quantity equal to the genetic variance.  This quantity in invariant in the sense that the many complexities of mating, environment, and genetic interactions are subsumed into a single value that does not depend on the large number of details that can differ.  Invariant quantities tell us what does not matter; what is left is all that matters.  Many of the deepest insights in science have arisen from a clear understanding of what matters and what does not matter---from a clear expression of invariance.

The second component of the theorem is an approximate conservation law.  The total change in fitness tends to remain close to zero, so the deterioration of the environment tends to be equal and opposite to the rate of increase in fitness caused by natural selection.  This conservation law captures the ever improving adaptation of individuals offset by the increasing pressure of competition from the improved adaptation of other individuals. Although other factors also contribute to the deterioration of the environment, Fisher emphasized this balance between individual improvement and enhanced competitive pressure.  

The theorem is clearly designed to express laws rather to than to calculate long term dynamics.  Laws play a key role in understanding natural phenomena.  Laws also set boundaries that must be satisfied by all systems---necessary but not sufficient conditions by which we may calculate the dynamics of systems.  To the extent that one wishes to calculate dynamics, the theorem is limited by its description of laws rather than dynamics. 

\citeA[p.~39]{fisher58the-genetical} used the principles of statistical mechanics to obtain simple laws: 

\begin{quote}
The regularity of [natural selection] is in fact guaranteed by the same circumstance which makes a statistical assemblage of particles, such as a bubble of gas obey, without appreciable deviation, the laws of gases.
\end{quote}

To understand statistical mechanics, think of each allele in a population as an individual particle, like a particular atom of an element.  The whole population is a collection of particles divided into discrete sets, each set forming the genotype of an individual.  The population of a large number of alleles divided into many distinct genotypes is like a large collection of atoms divided into many distinct molecules.  

One can study the dynamics of a collection in two distinct ways: particle dynamics or statistical mechanics.  In particle based dynamics, one analyzes the dynamics of the aggregate population by following the dynamics of each particle.  In genetics, that would mean studying the dynamics of the population and its fitness by analyzing the dynamics all alleles with respect to their assortment into genotypes.  Particle based dynamics is the most complete description possible.  It is also hopelessly complex for all but the most unrealistically reduced of systems.  Thus, any physical study of large aggregates applies statistical mechanics.

In statistical mechanics, one reduces all of the complex dynamics of the individual particles to a simple statistical summary.  For example, the movement of each particle can be thought of as fluctuation, and each fluctuation is typically influenced by interactions with many other particles.  To study explicit dynamics, each fluctuation of each particle must be analyzed with respect to all of the interactions between particles.  In genetics, we can think of a fluctuation as a change in fitness caused by a particular gene, each fitness fluctuation ascribed to a gene depending on the interaction of that gene with many other genes.  

To study statistical mechanics, we may use the variance of the individual fluctuations---a single aggregate measure that summarizes the overall intensity of fluctuation in the whole population.  Thus, the variance in fitness of the individual genes is the single aggregate measure of genetic variance in the population.  Fisher's fundamental theorem shows that a single aggregate measure of variance is sufficient to fix the total change in fitness caused by natural selection.  The reduction in complexity is almost magical.  Much of our understanding of the natural world arises from being able to reduce the overwhelming complexity of the dynamics of many particles to simple aggregate measures that capture essential features of system behavior.

\citeA[p.~39]{fisher58the-genetical} felt very strongly about the deep power of the statistical laws of nature and of what he accomplished with his theorem:
\begin{quote}
It will be noticed that the fundamental theorem proved above bears some remarkable resemblances to the second law of thermodynamics.  Both are properties of populations, or aggregates, true irrespective of the nature of the units which compose them; both are statistical laws; each requires the constant increase of a measurable quantity, in the one case the entropy of a physical system and in the other the fitness$\ldots$of a biological population$\ldots$Professor Eddington has recently remarked that `The law that entropy always increases---the second law of thermo\-dynamics---holds, I think, the supreme position among the laws of nature'. It is not a little instructive that so similar a law should hold the supreme position among the biological sciences.
\end{quote}

One of Fisher's main goals for his book was to demonstrate the law-like character of natural selection in shaping the biological world.  He wanted to put to rest many of the groundless criticisms of natural selection that we continue to hear today.  \citeA[p.~40]{fisher58the-genetical} continued:
\begin{quote}
The statement of the principle of Natural Selection in the form of a theorem determining the rate of progress of a species in fitness$\ldots$puts us in a position to judge of the validity of the objection which has been made, that the principle of Natural Selection depends on a succession of favourable chances.  The objection is more in the nature of an innuendo than of a criticism, for it depends for its force upon the ambiguity of the word chance, in its popular uses.  The income derived from a Casino by its proprietor may, in one sense, be said to depend upon a suggestion of improbability more appropriate to the hopes of the patrons of his establishment.  It is easy without any very profound logical analysis to perceive the difference between a succession of favourable deviations from the laws of chance, and on the other hand, the continuous and cumulative action of these laws.  It is on the latter that the principle of Natural Selection relies.
\end{quote}

These quotes help to understand Fisher's motivation with regard to the fundamental theorem and to analyze his various arguments with Wright about the fundamental theorem and the adaptive landscape.  Fisher viewed the fundamental theorem as an invariance law about natural selection rather than an expression of evolutionary dynamics.  He fully acknowledged that other evolutionary processes affected dynamics. The fundamental theorem is not a complete statement of evolutionary change, only a statement about natural selection: ``Natural Selection is not Evolution'' is the first sentence of Fisher's book \cite[p.~vii]{fisher58the-genetical}.

By contrast, Wright's mathematical theories analyzed gene frequency dynamics, that is, the full dynamics of the individual particles that make up the system.  Wright needed to study particle dynamics because he wanted to characterize those situations in which evolutionary systems change from being dominated by particular particle interactions through a transition to which alternative particle interactions dominate.  Put another way, Wright was concerned with epistatic gene interactions that bound a population to a local fitness peak, and the change in gene frequencies that would alter the gene combinations to shift a population to a different fitness peak.  To study mathematically that sort of peak shift, Wright did not study full dynamics of real systems, which is not possible, but instead reduced system size to a small number of genes (particles) to capture the particular interactions in an explicit way.  

\section*{Key points in the Fisher-Wright controversy}

My main argument is that Fisher and Wright talked past each other with regard to the fundamental theorem and the adaptive landscape.  They did so because each was usually arguing about some other issue, although in a way that often left the subtext obscure.  

There are four main ways in which Fisher and Wright talked past each other.  These four items help to parse the Fisher-Wright controversy and to understand more deeply the history and key concepts of evolutionary theory.

\subsection*{1. Wright lacked interest in Fisher's general laws}

Wright always tried to parse the fundamental theorem in relation to its consequences for long-term evolutionary dynamics.  I think almost everyone analyzed the fundamental theorem in this way.  The reason is that mathematical theory in science is often regarded as simple dynamical expressions: start with initial conditions and hypothesized rules of change and calculate the predicted outcome.  The predicted outcome is the dynamical expression of the future state of the system given the initial conditions and rules of change.  

Certainly all of Wright's mathematical theory is cast in this standard dynamical framework.  His mathematics may be technically dense at times, but the framing and goals are usually very clear in regard to the standard view of dynamical theory in science.  

Reading Fisher's exposition of the fundamental theorem in the context of his book, I find it hard to understand how everyone could have tried to force the theorem into this standard dynamical context.  As I showed above, Fisher gave an invariance law and an approximate conservation law.  The invariance is that the rate of change in fitness caused by natural selection is always the genetic variance in fitness.  He made clear that another component of evolutionary change in total fitness must always be ascribed to the deterioration of the environment. In the approximate conservation law, mean fitness remains nearly constant, thus the deterioration of the environment must usually be nearly equal and opposite to the increase by natural selection in order to maintain nearly constant total fitness. 

In Wright's \citeyear{wright77evolution} magnum opus, the index entry for ``Fundamental theorem of natural selection'' says ``\textit{See\/} Evolutionary transformation (panmictic species).''  It is hard to think of Wright as being ironical.  But the irony is certainly there: Fisher's \citeyear{fisher41average} scathing criticism of Wright's adaptive landscape focused most strongly on Wright's assumption of random mating (panmixia) in his early formulation, as opposed to Fisher's own carefully chosen definitions to distinguish the effects of genes under nonrandom mating.  Wright certainly felt the sting of Fisher's criticism to which he replied in \citeA{wright64stochastic}.  But Wright took many opportunities after 1941 to label Fisher's theorem as a statement about randomly breeding, panmictic populations.  For example, \citeA[p.~425]{wright77evolution}:
\begin{quote}
As noted, Fisher's theorem holds strictly only under the assumption of random combination of loci.  It applies in equilibrium populations with respect to genes with wholly independent effects, in spite of linkage.
\end{quote}

\subsection*{2. Long term dynamics and the rate of adaptation}

Wright labeled Fisher's theorem as one of random mating for two reasons.  First, Wright ignored Fisher's development of laws about natural selection and instead interpreted Fisher with respect to a theory of long term evolutionary dynamics.  Technically, this means that Wright ignored Fisher's partition of total evolutionary change into two components, natural selection and change of the environment.  In Fisher's theory, some of the total evolutionary change under nonrandom mating falls into the change of the environment, in the sense that changing genotype frequencies alter the genetic environment of each gene \cite{frank92fishers}.  

Second, Wright wanted to create a contrast between their views on the rate of adaptation. For example, in \citeA[p.~122]{wright88surfaces}:
\begin{quote}
Fisher's ``fundamental theorem of natural selection'' was concerned with the total combined effects of alleles at multiple loci under the assumption of panmixia in the species as a whole. He recognized that it was an exceedingly slow process.
\end{quote}

Fisher was interested in this debate about long term dynamics and the rate of adaptation. He realized that these issues did not concern his fundamental theorem about laws. So, when arguing with Wright about the rate of adaptation, he never answered directly with respect to the fundamental theorem.

Wright repeatedly stated that Fisher's fundamental theorem leads to a very slow rate of long term adaptation.  I have not found any statement by Fisher about the slowness of adaptive evolution following from either his theorem or his view of adaptation.  I believe Wright emphasized the slowness of Fisher's view, because Wright believed that in a large, mixed population, the only source of new variation for adaptation must come from new favorable mutations.  The rate of adaptation by new favorable mutations would, in Wright's view, be slow. Wright ascribed that view of slowness to Fisher, even though Fisher rejected such a conclusion.  

In the quotes given above from \citeA{fisher50the-sewall}, Fisher made clear that he believed fluctuating selection pressures are common in nature.  Under fluctuating selection, gene frequencies may be perturbed in ways that change the combinations of interacting genes favored by natural selection.  Once such changes occur, rapid adaptive change may follow by the process Wright, but not Fisher, ascribed to the fundamental theorem.  

In spite of the clear comments in \citeA{fisher50the-sewall}, Wright continued to claim that Fisher believed adaptive evolution to be an exceedingly slow process in large, mixed populations.  Wright contrasted this Fisherian strawman with his own view of rapid adaptive change driven by population subdivision, random perturbation of gene frequencies by small population size, followed by rapid adaptive evolution when new gene combinations are favored by natural selection.  \citeA{fisher50the-sewall} agreed that Wright's theory would lead to rapid adaptive evolution, but they regarded that theory as neither necessary nor likely for the explanation of rapid adaptive evolution.  

\subsection*{3. Additivity versus genetic interactions}

Wright's whole program turned on the novel variation generated by changes in the favored combinations of interacting genes.  Those changes in favored combinations do not depend on new mutations, but rather on fluctuations in gene frequencies.  For example, a particular gene cannot increase in frequency if it works well only with another gene that is rare and works poorly with the common alternative gene.  A fluctuation that makes the rare gene become common changes the situation, allowing the beneficial combination to be favored.  

Wright repeatedly characterized Fisher's theorem as incapable of dealing with such genetic interactions.  This characterization must have puzzled Fisher, who devoted much of his famous 1918 paper on quantitative genetics to the complexities of genetic interactions.  That paper describes the explicit partitioning of genetic variance into components that arise from the direct effects of each gene---the average effect---and the interactions that arise from dominance and epistasis.  In the partitioning of total genetic variance, one adds the direct effects of each gene, that additive component is often called the additive variance.  But that component does not arise physiologically from constant additive contributions of separate genes.  The direct effect of a gene depends on the frequencies of all other genes with which it interacts, and the direct effect changes with the changing frequencies of those interacting loci.  

It is true that inheritance is controlled by the sum of the direct effects calculated for each independent gene, because in the short term, it is only those direct effects that get transmitted from parent to offspring after the sexual mixture of parental genomes.  The brilliance of Fisher's analysis was to find a simple expression for the heritable component within the complex system of genetic interactions that he assumed was universal.  The fundamental theorem was a direct descendant of the statistical approach to genetic interactions originated in 1918. 

Why did Wright ascribe to Fisher the assumption of constant additive effects of separate genes?  To understand the shifting standard of the average effect in a theory that predicts the rate of change in mean fitness, one has to understand that the fundamental theorem is not about evolutionary dynamics but instead about the invariant quantity of genetic variance with respect to the natural selection component of evolutionary change.  Wright was not interested in that invariant law, but rather in his own world view with respect to long-term evolutionary dynamics.  Thus, he misrepresented the theorem, because he only discussed it within his own frame of reference.  

In addition, I think that Wright favored sharp distinctions between Fisher's work and his own: Fisher associated with additive, independent gene action in large randomly breeding populations versus Wright associated with complex genetic interactions in small subdivided populations.  By this characterization, Wright linked Fisher to slow adaptive change limited by the flow of rare beneficial mutations, in contrast with Wright's own claim for rapid evolutionary change by random fluctuations of gene frequencies creating newly favored beneficial combinations.  Wright did not care about Fisher's laws or his statistical partitioning of genetic effects.  He did care deeply about his own shifting balance theory based on newly favored beneficial combinations of genes.  

Wright originally formulated the shifting balance theory during the early 1930s. He spent the following fifty years refining that theory primarily through the mathematical exploration of gene frequency dynamics over his metaphor of the adaptive landscape, which describes a surface of mean fitness.  

\subsection*{4. Wright's expression of fitness surfaces in an adaptive landscape}

\citeA[p.~118]{wright88surfaces} stated in his final paper [check if truly final]:
\begin{quote}
The$\ldots$diagrams$\ldots$represent cases in which the population is assumed to be so large and its individuals so mobile that there can be no significant effects of accidents of sampling, giving rise to the panmixia assumed by Fisher (1930) to be characteristic of species in nature under similar environmental conditions throughout the range. This assumption was basic to the derivation of his ``fundamental theorem of natural selection'' (1930, p. 35)$\ldots$The effects of these four processes [by which populations climb local peaks of fitness] may be calculated by means of Fisher's ``fundamental theorem of natural selection.''
\end{quote}

When arguing with Wright about the rate of adaptation and long-term evolutionary dynamics, Fisher rarely answered directly with respect to Wright's misrepresentation of the fundamental theorem.  On the few occasions that Fisher commented on Wright's use of the theorem, Fisher emphasized various isolated issues. But Fisher never tried to explain the distinction between his goal to formulate laws and Wright's goal to understand the rate of adaptation.  Perhaps the reason Fisher did not defend his view was expressed by \citeA{haldane64a-defense}: ``Fisher$\ldots$preferred attack to defense.''

\citeA{fisher41average} first noted the failure of Wright's formulation to handle nonrandom mating compared with the clear way in which nonrandom mating is handled by the fundamental theorem.  Fisher (p.~377) then stated:
\begin{quote}
It is, I think, clear from Sewall Wright's allusions to the subject that he has never clearly grasped the difficulties of interpretation of such expressions as
\begin{equation*}
\frac{\dd\Wbar}{\dd p}
\end{equation*}
in which the numerator involves the average of $W$ for a number of different genotypes greatly exceeding the number of gene frequencies $p$ on which their frequencies are taken to depend. It is likely, therefore, that he does not share my reasons for putting a particular and well defined meaning upon the phrase `\textit{average effect\/} of a gene substitution'.
\end{quote}
Fisher's point is that, for two different alleles at a locus, there are three different genotypes.  Thus, for 1000 different loci, there are $3^{1000}$ different genotypes, which far exceeds the size of any population.  Thus, the notion of $\dd p$ for a change in gene frequency may not have much meaning with respect to mean fitness, because a discrete additional copy of an allele for a change $\dd p$ in frequency must often be added in a way that creates a novel genotype, the composition of which would be hard to predict even in a very large population.  The discreteness of genotypes and the rarity of many genotypes means that average fitness cannot reliably change in a smooth and regular way with smooth changes in gene frequencies.  

\citeA[p.~219]{wright64stochastic} acknowledged that the mathematical formulation of the adaptive landscape could be used only when there are a small number of loci or one makes very regular assumptions about genetic effects:
\begin{quote}
The summation in the formula for $\Wbar$ has, however, as many terms as there are kinds of genotypes, $3^{1000}$ for 1000 pairs of alleles. This, of course, points to a practical difficulty in calculating $\D q$ for more than two or three pairs of interacting factors, unless a regular model is postulated.
\end{quote}

I suggested earlier that Fisher was reluctant to argue about the fundamental theorem as a law rather than a statement of dynamics.  In Fisher's exchanges with Wright, Fisher usually kept to the issue of long-term evolutionary dynamics and the rate of adaptation, topics that from his point of view were not directly related to the theorem.  One clear exception shows the point, from \citeA[p.~290]{fisher58polymorphism}:
\begin{quote}
I have never, indeed, written about $\wbar$ and its relationships$\ldots$the existence of such a potential function [i.e. a function nondecreasing in time]$\ldots$is not a general property of natural populations, but arises only in the special and restricted cases which Wright has chosen to consider.

I should not have alluded to this storm in a tea-cup, but for the circumstance that I mean to put forward some ideas on$\ldots$the possible adaptive value of polymorphisms, and, incidentally, to express my personal opinion that Dobzhansky was right in regarding polymorphism as very often properly described as an adaptation to the conditions of life in which a species finds itself, but for reasons quite distinct from the direct action of Natural Selection, by which the polymorphism is maintained, or indeed from Natural Selection as it acts among the individuals of any one interbreeding population.
\end{quote}

Another brief mention by Fisher also points to the way in which Wright's formulation of change in mean fitness differed from Fisher's own view of natural selection: ``In regard to selection theory, objection should be  taken to Wright's equation [the expression $\dd\Wbar/\dd q$] principally because it represents natural selection, which in reality acts upon individuals, as though it were governed by the average condition of the species or inter-breeding group'' \cite[p.~58]{fisher41average}.

\citeA[p.~219]{wright64stochastic} responded to that criticism many years later:
\begin{quote}
As I understood it, Fisher [was]$\ldots$trying to arrive at a theorem on the rate of increase of ``fitness'' under natural selection that applies to a species as a whole.	My purpose was to obtain a formula for change of gene frequency in a random breeding deme in cases that involve factor interaction.
\end{quote}

I think it is generally true that Wright was interested in gene frequency change rather than mean fitness.  But his adaptive landscape metaphor of climbing fitness peaks by natural selection was a prominent part of his view.  Indeed, in \citeA[p.~241]{wright42statistical}, he makes clear that he considered how climbing adaptive peaks does directly affect the mean fitness of populations:
\begin{quote}
These [gene frequency] changes will be such that the mean selective value of the populations changes approximately by the amount
\begin{equation*}
\D\Wbar = \sum\left(\D q \partial\Wbar\big/\partial q \right)
\end{equation*}
the species moving up the steepest gradient in the surface $\Wbar$ except as affected by mutation pressures.
\end{quote}
This statement is clearly about the rate of change in the mean fitness of populations, contradicting Wright's later comment. Perhaps the Fisher-Wright controversy remains alive long after the combatants have passed because of the odd dissonance between what these two seemed to be saying at any point, what they had said previously, and the  underlying and often hidden basis of their disagreements.

\section*{Acknowledgments}

My research is supported by National Science Foundation grant EF-0822399, National Institute of General Medical Sciences MIDAS Program grant U01-GM-76499, and a grant from the James S.~McDonnell Foundation.  

\vfill\eject

\bibliography{ftnsRefs}
\bibliographystyle{apaciteJEB}

\vfill\eject

\section*{Appendix}

Because Fisher never gave a precise definition of fitness, there is no historical basis for ascribing any particular expression to Fisher himself.  Fisher's vagueness about fitness also made it difficult to understand what might be meant by a deterioration of the environment in relation to an exact mathematical theorem.  

My own view is that Price's \citeyear{price95the-nature} definition of fitness is the only one that provides both mathematical consistency of the theorem and logical unity to Fisher's vision \cite{frank95george,frank97the-price,frank09natural}.  Here is the particular definition of fitness that provides a simple mathematical basis for the fundamental theorem and related topics.  The following is taken from \citeA{frank09natural}.

The fitness of a type defines the frequency of that type after evolutionary change.  Thus, we write $q_j' = q_j (w_j/\wbar)$, where $w_j$ is the fitness of the $j$th type, and $\wbar = \sum q_jw_j$ is the average fitness.  We may use $j$ to classify by any kind of type, such as allele, genotype, or any other predictor of fitness.

Here, $w_j$ is proportional to the fraction of the second population that derives from (maps to) type $j$ in the first population.  One often thinks of the second population as the descendants and the first population as the ancestors, but any pair of populations can be used, separated by an instant in time, by discrete generations, or by some other scale of divergence that is not related at all to time.  The scale of divergence can be set by describing the point of measurement of the first population as $\Gth$, and the point of measurement of the second population as $\Gth'$.  Thus, $q'_j$ does not mean the fraction of the population at $\Gth'$ of type $j$, but rather the fraction of the population at $\Gth'$ that derives from type $j$ at $\Gth$. 

The fitness measure, $w$, can be thought of in terms of the number of progeny derived from each type.  In particular, let the number of individuals of type $j$ at $\Gth$ be $N_j=Nq_j$, where $N$ is the total size of the population.  Similarly, at $\Gth'$, let $N_j'=N'q_j'$.  Then $\wbar=N'/N$, and $w_j=N_j'/N_j$.

Fitness can alternatively be measured by the rate of change in numbers, sometimes called the Malthusian rate of increase, $m$.  This is the measure that Fisher typically used.  To obtain the Malthusian rate of increase with respect to an infinitesimal change in scale, $\GD\Gth\rightarrow\dd\Gth$, define the overdot as the differential $\dd/\dd\Gth$, and write
\begin{align}
		\frac{\dot{q}_j}{q_j} &= \dot{\log}(q_j)\notag\\
					&= \dot{\log}(N_j/N)\notag\\
					&= \dot{\log}(N_j)-\dot{\log}(N)\notag\\
					&= \dot{N}_j/N_j-\dot{N}/N\notag\\
					&= m_j - \bar{m}\notag\\
					&=a_j,\notag
\end{align}
where $a_j$ is the average excess in fitness.  Because the changes here are infinitesimal, corresponding to continuous time and in \Eq{aveExcess} to $\Wbar\rightarrow1$, the expression here is equivalent to the expression in \Eq{aveExcess} for the average excess in fitness.

\end{document}